\documentstyle[preprint,aps]{revtex}
\begin {document}
\draft
\preprint{UCI-TR 96-38}
\title{Sum Rule for Heavy Meson Decay Widths}
\author{Myron Bander\footnote{Electronic address:
mbander@funth.ps.uci.edu}\addtocounter{footnote}{3}%
}
\address{
Department of Physics, University of California, Irvine, California
92657-4575}
\date{\today}
\maketitle
\begin{abstract}
A sum rule relating the widths of the decays of mesons belonging to
heavy quark multiplets, having the same parity and light quark spin
$j$, into the low lying $0^-$ and $1^-$ multiplet is obtained. As
this sum rule follows from properties of the axial charges, it is
protected from heavy quark spin dependent $1/m_Q$ corrections. The
sum rule works well for mesons containing a heavy charmed quark and,
surprisingly, for resonances containing a strange quark. 
\end{abstract}

\pacs{PACS numbers: 12.39.Hg,13.25.Ft,13.25.Es}
The heavy quark effective theory, recently reviewed in
\cite{neubert,falk1,falk2}, has been applied, with varied success,
to the strong decays of meson resonances containing one such quark. For
example, the theoretical value for the ratio
\begin{equation}\label{ratio1}
\frac{\Gamma\left(D_2^*(2460)\to D+\pi\right )}
   {\Gamma\left(D_2^*(2460)\right )\to D^*+\pi)}=2.3
\end{equation}
is in good agreement with the experimental result $2.3\pm
0.6$ or $1.9\pm 1.1$ \cite{exper1} (depending on the charge of the
resonance). However, the theoretically obtained ratio
\begin{equation}\label{ratio2}
\frac{\Gamma\left(D_1^*(2420)\right )\to D^*+\pi)}
  {\Gamma\left(D_2^*(2460)\right )\to D+\pi)+\Gamma\left(D_2^*(2460)
     \right )\to D^*+\pi)}=0.3 
\end{equation}
is in disagreement with the experimentally determined one \cite{exper2} 
of $0.82\pm 0.23$. Eq.~(\ref{ratio1}) gets no correction to order
$1/m_Q$, ($m_Q$ is the heavy quark mass) whereas the one of
eq.~(\ref{ratio2}) does \cite{falk2}.

In this work we will present a sum rule relating widths of resonances
decaying to $D$ and to $D^*$ (and their analogs for other heavy quark
mesons); there results are robust to $1/m_Q$ corrections.  As we shall
see, not only are these sum rules satisfied for resonances containing a
charmed quark, but, surprisingly, work well for $K$ resonances. 

The results we shall find are valid in the soft pion limit and, in
part, are derived in a manner similar to the one used to obtain the
Adler-Weisberger relation \cite{AW}. With $Q^{\pm}_5$, the isospin
plus/minus component of the axial charge, we consider
\begin{equation}\label{Qprod}
  I_H(p;\Delta )\delta^3(p-p')=\sum_{m_n^2\varepsilon\Delta}
   \left (\langle H, p|Q^-_5|n\rangle\langle n|Q^+_5|H, p'\rangle
     -\langle H, p|Q^+_5|n\rangle\langle n|Q^-_5|H, p'\rangle
       \right )\, ;
\end{equation}  
$|H,p\rangle$ is any state with momentum $p$ and the summation is
restricted to states whose mass squared lies in an interval
$\Delta$. Using the techniques developed in Ref.~\cite{FFR} we may
show that
\begin{equation}\label{partialaw}
  I_H(p;\Delta )=\frac{4f_\pi^2}{\pi}\int_{\Delta}\frac{ds}{(s-m_H^2)^2}
   \mbox{\rm Im}\,\left ( {\cal T}^{(\pi^-H)}(s)
    -{\cal T}^{(\pi^+H)}(s)\right )\, ;
\end{equation}
in the above, $f_\pi=93$ MeV is the pion decay constant, ${\cal
T}^{(\pi^{\pm}H)}(s)$ is the amplitude, at center of mass energy squared 
$s$, for $\pi^{\pm}H$ scattering and the integration over $s$ is
restricted to the interval $\Delta$. (Relaxing this restriction yields
the Adler-Weisberger sum rule.)  

We shall apply eq.~(\ref{partialaw}) in cases where the state $H$ is
the low lying $0^-$ or $1^-$ meson containing one heavy quark, for
definiteness let us say the $D$ and $D^*$. As in the heavy quark limit
interactions of pions with these mesons is independent of the spin of
the heavy quark we obtain
\begin{equation}\label{sumrule1}
  \int_{\Delta}\frac{ds}{(s-m_D^2)^2}\mbox{\rm Im}\,
   \left ( {\cal T}^{(\pi^-D)}(s)-{\cal T}^{(\pi^+D)}(s)\right )=
    \int_{\Delta}\frac{ds}{(s-m_{D^*}^2)^2}\mbox{\rm Im}\,
     \left ({\cal T}^{(\pi^-D^*)}(s)-{\cal T}^{(\pi^+D^*)}(s)\right )\, ;
\end{equation}
${\cal T}^{(\pi^{\pm}D^*)}$ is the $\pi^{\pm}D^*$ amplitude for any $D^*$ spin
or helicity state. Eq.~(\ref{sumrule1}) is obviously valid in the
$m_Q\to\infty$ limit and we didn't need the previous discussion to obtain
it; we shall show that as as a result of its connection to the axial
charge it survives $1/m_Q$ corrections that depend on the heavy quark
spin. As there are no isospin 3/2 resonances, in a resonance
approximation this relation this becomes
\begin{equation}\label{sumrule2}
  \sum_{R\varepsilon\Delta} (2j_r+1)\frac{\Gamma(R\to D\pi)}{p^3}
    =\frac{1}{3}\sum_{R\varepsilon\Delta} (2j_R+1)
       \frac{\Gamma(R\to D^*\pi)}{p^3}\, ;
\end{equation}
the summation extends over resonances whose mass squared is in the
interval $\Delta$ and $p$ is the decay momentum. As eq.~(\ref{sumrule1})
involves a difference of amplitudes, nonresonant scattering is expected
to cancel and eq.~(\ref{sumrule2}) should be a good approximation.

In order to show that there are no $1/m_Q$ corrections depending on
the heavy quark spin and for other further discussions it is useful to
review some of the heavy quark spectroscopy and effective field
theory. The heavy quark multiplets, both resonant and non-resonant,
are labeled by the total angular momentum, $j=l+s$, of the light
quark; this in turn is coupled to the spin-1/2 heavy quark. For $l=0,\
j=1/2$ resonant states are $D$ and $D^*$ with spin-parity $0^-$ and
$1^-$ respectively; for $l=1,\ j=1/2$ we have a $D_{0^+}$ and a
${D_{1^+}}'$, while for $l=1,\ j=3/2$ we find a second $D_{1^+}$ (the
reason for the prime in the notation for the previous state) and a
$D_{2^+}$. The latter may be identified with the $D_1(2420)$ and
$D_2(2460)$ respectively, while the former has not yet been seen. It
is convenient to introduce fixed velocity fields
\cite{neubert,falk1,falk2}, which for $\mbox{\bf v}=0$ are $2\times 2$
matrices.
\begin{eqnarray}\label{fixedvfields}
 H&=&\frac{1}{\sqrt 2}\left (D+\sigma_aD^{*a}\right )\, ,\nonumber\\
  S&=&\frac{1}{\sqrt 2}\left (D_{0^+}-\sigma_a{D^a_{1^+}}'\right )\, ,\\
   T^a&=&\frac{1}{\sqrt 2}\left [\sqrt{\frac{2}{3}}D^a_{1^+}+
    \left( D^{ab}_{2^+}-i\sqrt{\frac{1}{6}}\epsilon^{abc}D^c_{1^+}
     \right )\sigma_b\right ]\, .\nonumber
\end{eqnarray}
The axial charges $Q_5^i$ connect ground state the $l=0,\ j=1/2$ 
multiplet with all $l=1,\ j=1/2$ multiplets. 
\begin{equation}\label{Q51}
  Q_5^i=\sum_r g_r \mbox{\rm Tr}H\frac{\tau^i}{2} S^{(r)\dag} +
    \mbox{\rm h.c.}\, ; 
\end{equation}
the summation is over all $l=0,\ j=1/2$ multiplets and the $g_r$ are 
constants determining the strengths of the matrix element of the axial
charge between the states in question. Using the fixed velocity field
equal time commutation relations
\begin{eqnarray}\label{commrels}
  \left[H_{ab},\, H_{cd}^{\dag}\right ]&=&i\delta_{ad}\delta_{bc}\, ,
    \nonumber\\
  \left[S^{(r)}_{ab},\, S_{cd}^{(s)\dag}\right ]&=&
    i\delta^{rs}\delta_{ad}\delta_{bc} \, .
\end{eqnarray} 
the current algebra relations 
\begin{equation}\label{curtalg}
\left[Q_5^i,\, Q_5^j\right ]=if^{ijk}Q^k\, ,
\end{equation}
with $Q^k$ the vector isospin charge, imply $\sum_r g_r^2=1$,
independent of the heavy quark mass. Corrections to the axial charge
matrix elements that are not invariant under heavy spin rotations, such
as 
\begin{equation}\label{deltaQ5}
  \delta Q_5^i=\sum_r \frac{h_r}{m_Q} \mbox{\rm Tr}\sigma^\alpha 
       H\frac{\tau^i}{2}\sigma_\alpha S^{(r)\dag} + \mbox{\rm h.c.}\, ,
\end{equation}
are not allowed as these could not be accommodated by current algebra;
due to this eq.~(\ref{sumrule2}) has no heavy quark spin dependent
$1/m_Q$ corrections. For an interval $\Delta$ dominated by one
heavy quark multiplet with light quark angular momentum $j$ we obtain
\begin{equation}\label{result}
  \frac{1}{3}\sum_{S_R=j\pm 1/2} (2S_R+1)
    \frac{\Gamma(R\to D^*\pi)}{p^3}=
  \sum_{S_R=j\pm 1/2} (2S_R+1)\frac{\Gamma(R\to D\pi)}{p^3}\, ;
\end{equation}
this is the main result of this paper.

For the charmed heavy quark the $D_{1^+}(2420)$ and the
$D_{2^+}(2460)$ form such a multiplet with $j=3/2$. Table
\ref{result1} contains the experimental data and the last column gives
the numerical values for the left and right hand sides of
eq.~(\ref{result}); this sum rule is very well satisfied. We note that
this is an D wave decay and the widths are proportional to $p^5$; yet
with a broad range of decay momenta it is the $\Gamma/p^3$'s that are
related.

There are no data on other $D$ multiplets to compare with
eq.~(\ref{result}), however it is tempting to look at the $K$ meson
system, even though the mass of the strange quark is generally taken
to be to light for the heavy quark formalism to apply. Results for
this system analogous to the one shown in Table \ref{result1} are
presented in Table \ref{result2}; again the agreement is very good. In
the $K$ system data exist for the $l=1,\ j=1/2$ resonances, namely
$K_{1^+}(1400)$ and $K_{0^+}(1430)$. The result shown in Table
\ref{result3}, although not as spectacular as the previous two, is
reasonable.

It is worthwhile to look at possible sources of corrections to
eq.~(\ref{result}) when applied to the $l=1,\ j=1/2\ K$ multiplet.  In
this analysis we have paired one of the $1^+$ resonances, the
$K_{1^+}(1270)$, with the $K_{2^+}(1430)$ and the other one, the
$K_{1^+}(1400)$ with the $K_{0^+}(1430)$. It is not clear whether it
should not have been the other way around, or more likely some linear
combination. A partial wave analysis \cite{Daum} indicates that the
$K_{1^+}(1400)$ decays to $K^*\pi$ almost exclusively in an S-wave and
is thus correctly paired. The $K_{1^+}(1270)$ has, in addition to a
D-wave component, an S-wave one. It is unclear how to include mixing
of the two $1^+$ states. The $K$ system points out a major source of
corrections to eq.~(\ref{result}). The resonance approximation,
eq.~(\ref{sumrule2}) should hold as long as the interval $\Delta$ is
dominated by the resonances in question. We have applied it,
eq.~(\ref{result}), to the case where there is only one multiplet in
$\Delta$; this should be valid for heavy meson systems. For the $K$
system the spin dependent splitting of the multiplets is large enough
that states belonging to other multiplets lie close by or even
intervene, e.g. the $K_{1^-}(1410)$ and the $K_{0^-}(1460)$. Testing
eq.~(\ref{sumrule2}) for the $K$ system by summing over several heavy
quark multiplets would be desirable but the necessary data on the
partial widths do not exist.

Eq.~(\ref{result}) should work well for other $D$ multiplets and even
better for the $B$ states. 

\nobreak 

\begin{table} 
\caption{$l=2$ D-Meson Decays. Column 2 gives the decay momentum for
the reaction in column 1 and the corresponding partial width is in
column 3. This partial width, multiplied by the initial spin
multiplicity and divided by the final state one, as in eq.~(12) are in
column 4 and the sum pertaining to a definite final state is shown in
the last column; the sum rule in eq.~(12), requires an equality of the
numbers in the last column.}
\begin{tabular}{ccccc}
{} & p (GeV) & $\Gamma$ (MeV) &
$\frac{(2S_i+1)\Gamma}{(2S_f+1)p^3}$ (GeV${}^{-2}$) &  Sum\\[2mm] 
\tableline
$D_{1^+}(2420)\to D^*\pi$ & $0.355$ & $18.9\pm 4.0$\tablenotemark[1] 
& $0.42\pm 0.11$ &{}\\[-5mm]  
{} & {} & {} & {} &$0.62\pm 0.12$\\[-5mm]
$D_{2^+}(2460)\to D^*\pi$ & $0.387$ & $7.0\pm 1.8$\tablenotemark[2] 
& $0.20\pm 0.05$ & {}\\
$D_{2^+}(2460)\to D\pi$& $0.503$ & $16.0\pm 1.8$\tablenotemark[2] 
&$0.67\pm 0.17$ & $0.67 \pm 0.17$\\ 
\end{tabular}\label{result1}
\tablenotetext[1]{Ref.\ \cite{exper1}, p.469.}
\tablenotetext[2]{Ref.\ \cite{exper1}, p.470.}
\end{table}

\begin{table} 
\caption{$l=2$ K-Meson Decays. Presentation is as in Table I.} 
\begin{tabular}{ccccc}
{}& p (GeV) & $\Gamma$ (MeV) &
$\frac{(2S_i+1)\Gamma}{(2S_f+1)p^3}$  (GeV${}^{-2}$)  &  Sum \\[2mm]
\tableline
$K_{1^+}(1270)\to K^*\pi$ & $0.301$ & $13.9\pm 4.2$\tablenotemark[1] 
& $0.51\pm 0.13$ & {}\\[-5mm] 
{} & {} & {} &{} & $1.06\pm 0.14$\\[-5mm]
$K_{2^+}(1430)\to K^*\pi$ & $0.423$ & $24.8\pm 1.7$\tablenotemark[2] 
& $0.55\pm 0.06$ & {}\\
$K_{2^+}(1430)\to K\pi$& $0.622$ & $48.9\pm 1.2$\tablenotemark[2] 
& $1.02\pm 0.11$ & $1.02 \pm 0.11$\\ 
\end{tabular}\label{result2}
\tablenotetext[1]{Ref.\ \cite{exper1}, p.432.}
\tablenotetext[2]{Ref.\ \cite{exper1}, p.434.}
\end{table}

\begin{table} 
\caption{$l=0$ K-Meson Decays. Presentation is as in Table I
except that, as each resonance has only one final state
no summation is needed; the sum rule in eq.~(12) requires an
equality of the numbers in the last column.} 
\begin{tabular}{cccc}
{} & p (GeV) & $\Gamma$ (MeV) &
$\frac{(2S_i+1)\Gamma}{(2S_f+1)p^3}$  (GeV${}^{-2}$)  \\[2mm]
\tableline
$K_{1^+}(1400)\to K^*\pi$ & $0.401$ & $117\pm 10$\tablenotemark[1] 
& $1.81\pm 0.17$ \\
$K_{0^+}(1430)\to K\pi$& $0.621$ & $267\pm 29$\tablenotemark[2] 
& $1.15\pm 0.13$ \\ 
\end{tabular}\label{result3}
\tablenotetext[1]{Ref.\ \cite{exper1}, p.433.}
\tablenotetext[2]{Ref.\ \cite{exper1}, p.434.}
\end{table} 


\begin{references} 
\bibitem{neubert}
M. Neubert, [hep-ph/9610385], to appear in the {\it Proceedings of the
Twentieth Johns 
Hopkins Workshop on Current Problems in Particle Theory, Heidelberg,
Germany, June 27-29, 1996}. 
\bibitem{falk1}
A.F. Falk, [hep-ph/9609380], {\it ibid.}.
\bibitem{falk2}
A.F. Falk, [hep-ph/9610363], to appear in the {\it Proceedings
of the XXIV\/th SLAC Summer Institute on Particle Physics, Stanford,
California, August 19-30, 1996}.
\bibitem{exper1}
R.M. Barnet {\it et al\/}, Phys.\ Rev.\ D{\bf 54}, 470 (1996).
\bibitem{exper2}
Ref. \cite{exper1}, p. 469.
\bibitem{AW}
W.I. Weisberger, Phys.\ Rev.\ Lett.\ {\bf 14}, 1047 (1965);
S.L. Adler, {\it ibid.}, {\bf 14}, 1051 (1965).
\bibitem{FFR}
S. Fubini, G. Furlan and C. Rossetti, {\it Nuovo Cimento} {\bf 40},
1171 (1965).  
\bibitem{Daum}
C. Daum {\it et al\/.}, Nucl. Phys. {\bf B}187, 1 (1981).
\end{references}
\end{document}